\documentclass[fleqn,10pt]{wlscirep}
\usepackage[utf8]{inputenc}
\usepackage[T1]{fontenc}
\usepackage{amsmath,amssymb,amsfonts}
\usepackage{algorithmic}
\usepackage{graphicx}
\usepackage{subcaption}
\usepackage{array}
\usepackage{algorithm,algorithmic}
\usepackage{hyperref}

\title{TFT-multi: simultaneous forecasting of vital sign trajectories in the ICU}

\author[1,2]{Rosemary Y. He}
\author[2,3,*]{Jeffrey N. Chiang}
\affil[1]{Department of Computer Science, University of California Los Angeles, CA 90095 USA}
\affil[2]{Department of Computational Medicine, University of California Los Angeles, CA 90095 USA}
\affil[3]{Department of Neurosurgery, University of California Los Angeles, CA 90095 USA}

\affil[*]{njchiang@g.ucla.edu}

\keywords{Multiple trajectory predictions, generative AI, time series, clinical support}

\begin{abstract}
Individual health trajectory forecasting is a major opportunity for computational methods to integrate with precision healthcare.
Recently developed generative AI models have demonstrated promising results in capturing short and long range dependencies in time series data. While these models have also been applied in healthcare, most of them only predict one feature at a time, which is unrealistic in a clinical setting where multiple measures are taken at once. In this work, we extend the framework temporal fusion transformer (TFT), a multi-horizon time series prediction tool, and propose TFT-multi, an end-to-end framework that can predict multiple vital trajectories simultaneously. We apply TFT-multi to forecast 5 vital signs recorded in the intensive care unit: blood pressure, pulse, SpO2, temperature and respiratory rate. We hypothesize that by jointly predicting these measures, which are often correlated with one another, we can make more accurate predictions, especially in variables with large missingness. We validate our model on the public MIMIC dataset and an independent institutional dataset, and demonstrate that this approach outperforms state-of-the-art univariate prediction tools including the original TFT and Prophet, as well as vector regression modeling for multivariate prediction. Furthermore, we perform a study case analysis by applying our pipeline to forecast blood pressure changes in response to actual and hypothetical pressor administration. 

\textbf{Keywords: Multiple trajectory predictions, generative AI, time series, clinical support}
\end{abstract}

\begin{document}
\flushbottom
\maketitle
\thispagestyle{empty}

\section*{Introduction}

In the age of big data, electronic health records (EHR) have become paramount in both clinical practice and research\cite{miriovsky2012importance}. One subset of EHR that has been of interest to both clinicians and researchers is time series data, often measured in laboratory values and vital signs, that can be applied to track disease trajectory from every few minutes to the span of years\cite{ruan2019representation}. Compared to cross-sectional (static) data, which remains constant, time series offer more information on patient development over time. These data can be leveraged to build trajectory prediction tools that can be used for surveillance (e.g., early warning systems\cite{ye2019real,zheng2020development}), and decision support (e.g., forecasting event of interest\cite{jawadi2024predicting} or response to interventions\cite{chen2019deep,cheng2021robust}). 
In this work, we develop a single approach that simultaneously predicts vital signs commonly recorded in the intensive care unit. We demonstrate that not only does this approach outperform state-of-the-art forecasting models, but it also can be used to forecast patient trajectories under hypothetical scenarios. 


With successful applications of generative artificial intelligence (genAI) models on time series data in finance\cite{takahashi2019modeling, vuletic2024fin} and climate science\cite{gao2022multi,huang2022time}, researchers have turned to AI-based tools to predict trajectories over time in healthcare\cite{allam2021analyzing,pham2017predicting}. Compared to traditional autoregressive methods such as the ARIMA model\cite{newbold1983arima}, AI-based models have the potential to better capture complex relationships between variables. In univariate modeling, Prophet\cite{taylor2018forecasting} is a state-of-the-art method to predict time series data in many applications. Another model that has been effective in capturing both long and short term trends is temporal fusion transformer (TFT)\cite{lim2021temporal}. In previous works, both methods have been used for vital sign forecasting in cardiovascular diseases\cite{ahmed2023multivariate} and sepsis patients\cite{bhatti2024vital}. However, the feasibility of implementing these approaches is limited by the fact that a separate model must be used for each outcome. Furthermore, vital signs may be informative of each other and their trajectories should not be predicted independently, prompting the need for models that can predict multiple variables simultaneously. The state-of-the-art for multivariate prediction models is the vector autoregressive (VAR) model\cite{zivot2006vector}, which captures the linear interdependencies among multiple time series. In this work, we extend the TFT model from univariate to multivariate prediction and develop an end-to-end pipeline that can simultaneously predict 5 common vital signs recorded in the intensive care unit (ICU). We demonstrate our method outperforms state-of-the-art methods including Prophet, the original TFT and VAR in terms of predictive power. In the future, our pipeline can be extended to real-time predictions and applied as an early warning system for patients in the ICU. In addition, we demonstrate a study case to apply our model for treatment effect estimation and decision support, generating hypothetical blood pressure trajectories with and without the administration of pressors.

We list our contributions in this work. First, we extend TFT from univariate to multivariate modeling to better reflect the clinical setting. To do so, we modify the input-output structure and the loss function to simultaneously predict multiple outcomes at 15-min intervals. Compared to univariate prediction models, our pipeline offers two main advantages: 1. it achieves better predictive performance by capturing relations between target variables; 2. it requires less computational resources as it can predict multi-horizon trajectories for multiple features all in one run. Second, we develop a preprocessing pipeline and use a masking technique during training to account for losses in the non-imputed values only. This allows our pipeline to learn dependencies across real values only and achieve strong performance in validation even in data with large sampling heterogeneity, which is common in time series healthcare data. Third, we propose an additional application for our pipeline in treatment effect estimation. We use the model to forecast individual responses to pressors, demonstrating the ability to compare alternative trajectories in hypothetical scenarios to supplement clinical decision support.

\section*{Results}

\subsection*{Multi-horizon prediction}
First, we compare performance in predicting 5 standard vital values in the ICU: mean arterial blood pressure (mean BP), pulse, temperature (temp), respiratory rate (resp) and SpO2 across four methods: Prophet\cite{taylor2018forecasting}, VAR\cite{zivot2006vector}, the original TFT\cite{lim2021temporal} and our method, TFT-multi. Because Prophet\cite{taylor2018forecasting} is a univariate forecasting model, five instances are trained to predict each of the vital sign time series. We note here that since Prophet\cite{taylor2018forecasting} requires the covariate inputs to also be known into the future, we only add age as a covariate. For TFT\cite{lim2021temporal}, we also train 5 instances with the same covariate inputs as our model to predict each of the vital signs. For VAR\cite{zivot2006vector}, we train a model to take in past vital sign trajectories and no exogenous variables, in accordance with convention. We note here that respiratory rate and temperature did not pass the Granger's causality test\cite{shojaie2022granger}, which determines if one time series can predict another time series. Since VAR assumes all variables to be predictive of one another\cite{zivot2006vector}, we did not include them in the model. By the same logic, we did not include temperature in the external validation set. For each model, we calculate the mean absolute error (MAE) in both validation datasets for all vital predictions against their true values. We examine only the time points with recorded and not imputed values, calculated as follows for each vital sign:
\begin{equation}
MAE = \frac{\sum_{y_i \in \Omega} \sum_{t = 1}^{t_{\text{max}}} |y_{it} - \hat{y}_{it}| \cdot \mathbb{I}_t({\text{real value}})}{\sum_{y_i \in \Omega} \sum_{t = 1}^{t_{\text{max}}} \mathbb{I}_t({\text{real value}})}
\end{equation}
In Table \ref{MAE}, we compare performances for each vital sign, with the best performing model in bold. The top 4 rows are calculated with the held-out MIMIC test set, while the bottom 4 with the external validation (EV) set. For TFT and TFT-multi, we take the 50th percentile prediction as the prediction. In the held-out test set, TFT-multi archives the lowest MAE in 3 out of 5 vitals, and is a close second to the best performing model for SpO2 (Prophet) and respiratory rate (TFT). In the external set, TFT-multi achieves the lowest MAE for 4 out of 5 vitals, and performs comparably with Prophet in SpO2, which Prophet is expected to perform well in as it is highly regular and cyclical. Compared to the original TFT model, there is a significant improvement in predicting vitals that have more missingness, such as SpO2.

\begin{table}[ht]
\centering
\begin{tabular}{|>{\centering\arraybackslash}m{3cm}|>{\centering\arraybackslash}m{2cm}|>{\centering\arraybackslash}m{2cm}|>{\centering\arraybackslash}m{2cm}|>{\centering\arraybackslash}m{2cm}|>{\centering\arraybackslash}m{2cm}|}
\hline
 & Mean BP & Pulse & SpO2 & Resp & Temp \\
 \hline
Prophet (test) & 13.4 [15.6] & 10.9 [11.2] & \textbf{1.7 [1.9]} & 5.6 [16.2] & 3.6 [3.9] \\
\hline
VAR & $>$100 [$>$1] & $>$100 [$>$1] & $>$100 [$>$1] & N/A & N/A \\
\hline
TFT & 11.9 [12] & 13.4 [13.2] & 14.4 [15] & \textbf{4 [12.6]} & 1.3 [1.3] \\
\hline
TFT-multi (ours) & \textbf{7.4 [8.1]} & \textbf{9 [8.9]} & 1.8 [2] & 4.5 [13] & \textbf{0.7 [1]} \\
\hline
\hline
Prophet (EV) & 8.2 [11] & 9.3 [11] & \textbf{1.7 [2.8]} & 3 [41] & 0.9 [0.9] \\
\hline
VAR & 6.5 [9.7] & 7.2 [8.6] & $>$100 [$>$1] & 2.5 [51.4] & N/A \\
\hline
TFT & 10.1 [12.4] & 8.7 [10.1] & 17.09 [22.8] & 2.42 [32.2] & 1.4 [1.4] \\
\hline
TFT-multi (ours) & \textbf{6.2 [8.9]} & \textbf{7.2 [7.8]} & 1.8 [2.8] & \textbf{2.4 [30.1]} & \textbf{0.8 [0.7]} \\
\hline
\end{tabular}
\caption{Model performance across five forecasted vital signs. Values indicate mean absolute error [mean absolute percentage error(\%)] for the held-out test and external validation sets. TFT-multi achieves the best performance (highlighted in bold) for 3 in test and 4 in external validation out of 5 features.}
\label{MAE}
\end{table}

\begin{figure*}[!htbp]
\centering
\subfloat[]{\includegraphics[width=0.33\textwidth]{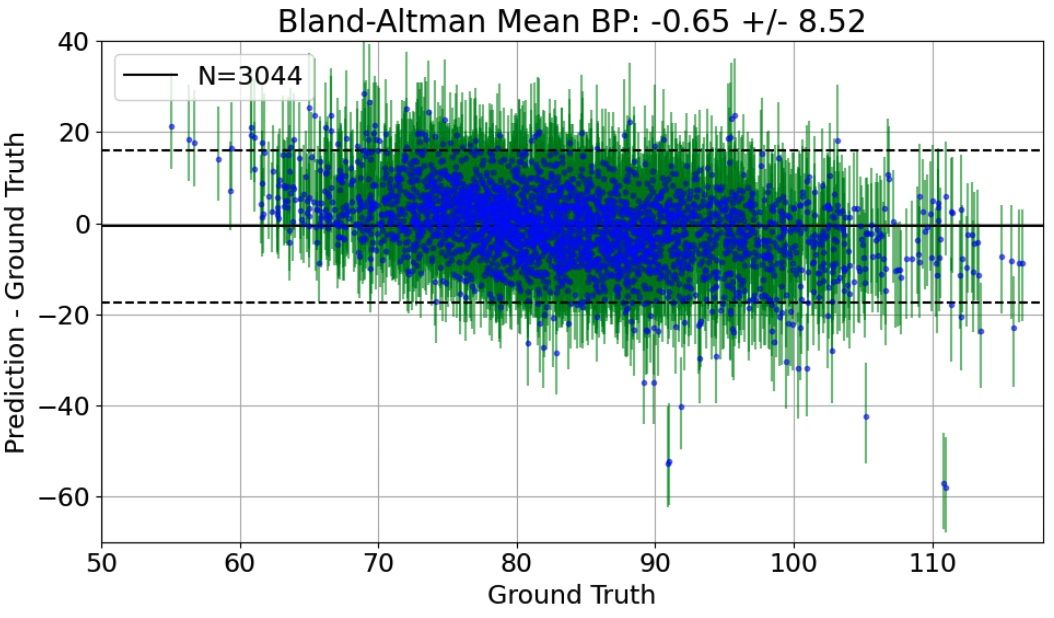}}\hfill
\subfloat[]{\includegraphics[width=0.34\textwidth]{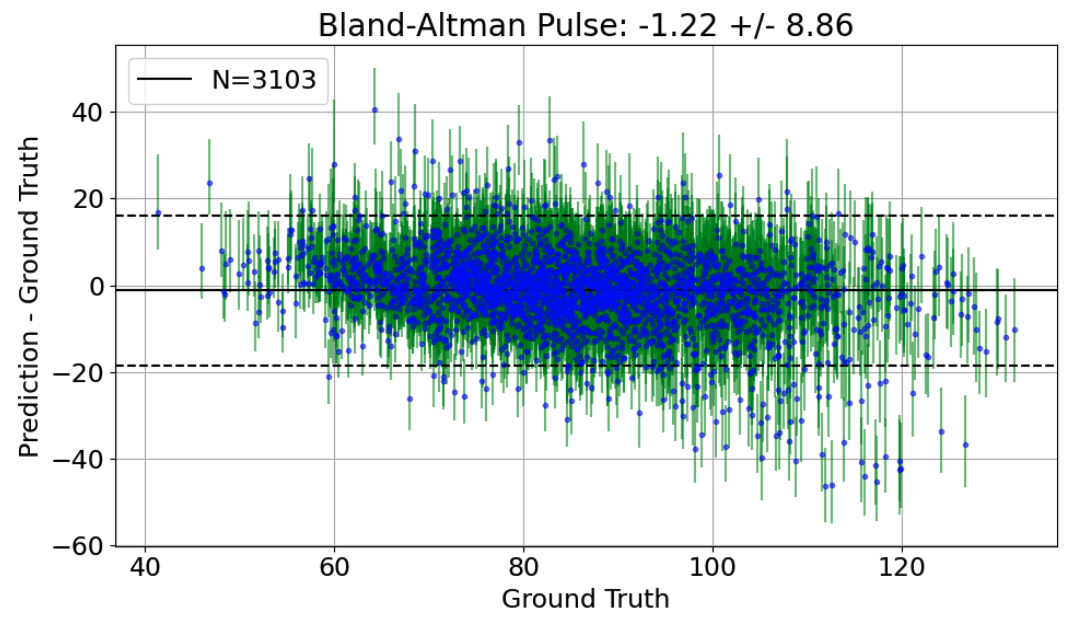}}\hfill
\subfloat[]{\includegraphics[width=0.33\textwidth]{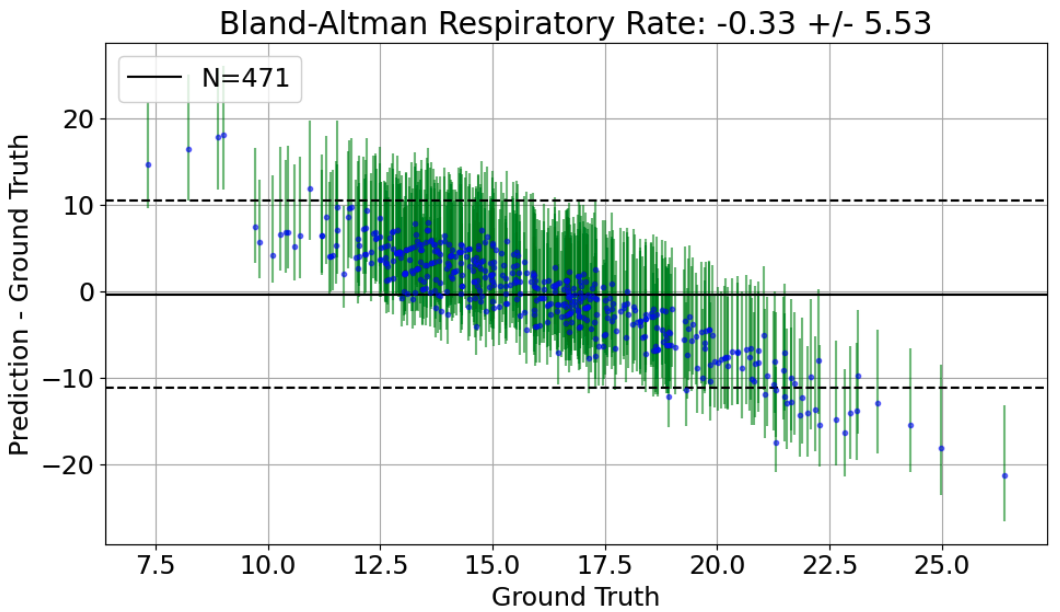}\label{bland_resp}}

\subfloat[]{\includegraphics[width=0.33\textwidth]{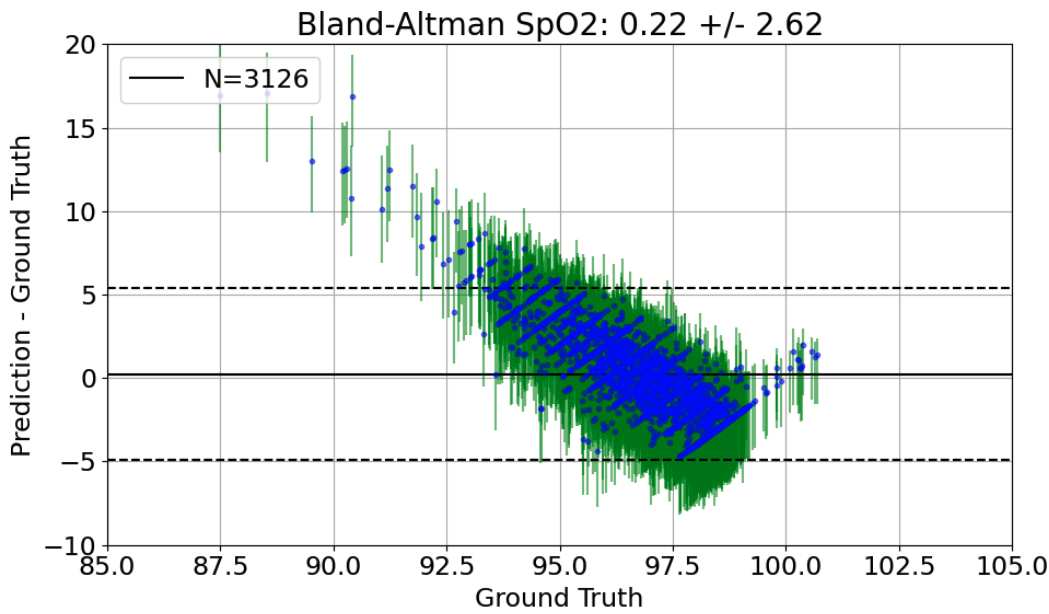}\label{bland_SpO2}}%
\hspace*{0.005\textwidth}%
\subfloat[]{\includegraphics[width=0.34\textwidth]{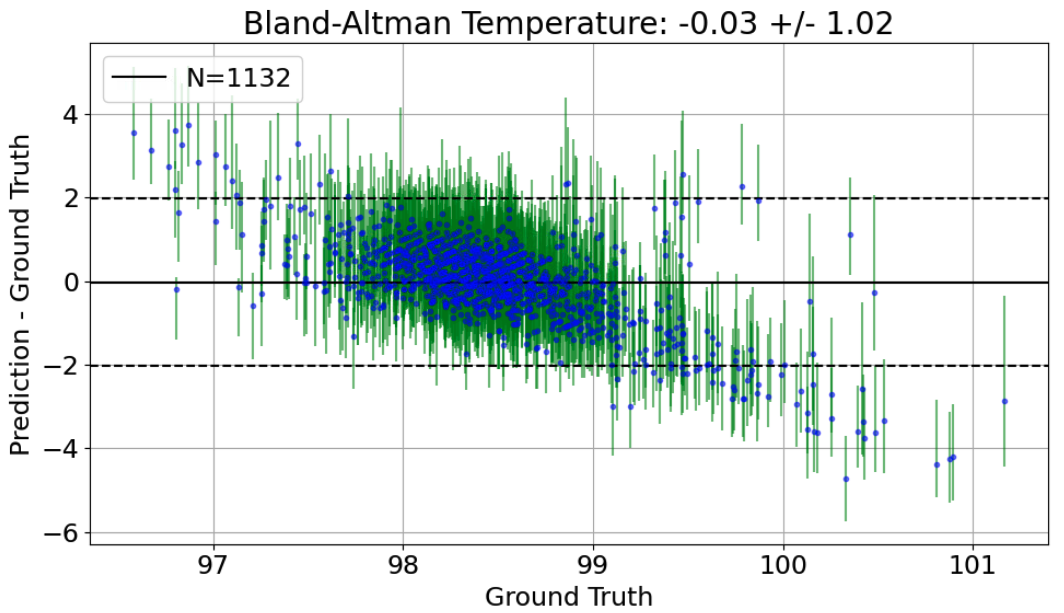}\label{bland_temp}}
\caption{Bland-Altman Plots for predictions in the held-out test set across vitals. Each point represents an observed time point. Vertical bars represent the upper and lower prediction bound errors; solid line indicate the mean difference, dashed lines indicate the mean difference $\pm1.96$ times the standard deviation of differences. TFT-multi is well calibrated in mean BP and pulse, but limited in features with more missingness.}\label{bland}
\end{figure*}

While MAE summarizes overall model performance, in diagnostic instruments the metric may be biased by the fact that the majority of measurements tend to fall within the ``normal'' range. In Fig. \ref{bland}, we examine the calibration of TFT-multi's prediction at the 50th percentile with the ground truth in the test set across all vitals using a Bland-Altman plot\cite{altman1983measurement,giavarina2015understanding}. While TFT-multi appears to be well calibrated for mean BP and Pulse, we observe limited calibration for respiratory rate, SpO2, and temperature despite low MAE values.  

\begin{figure}[!t]
\centerline{\includegraphics[width=0.7\columnwidth]{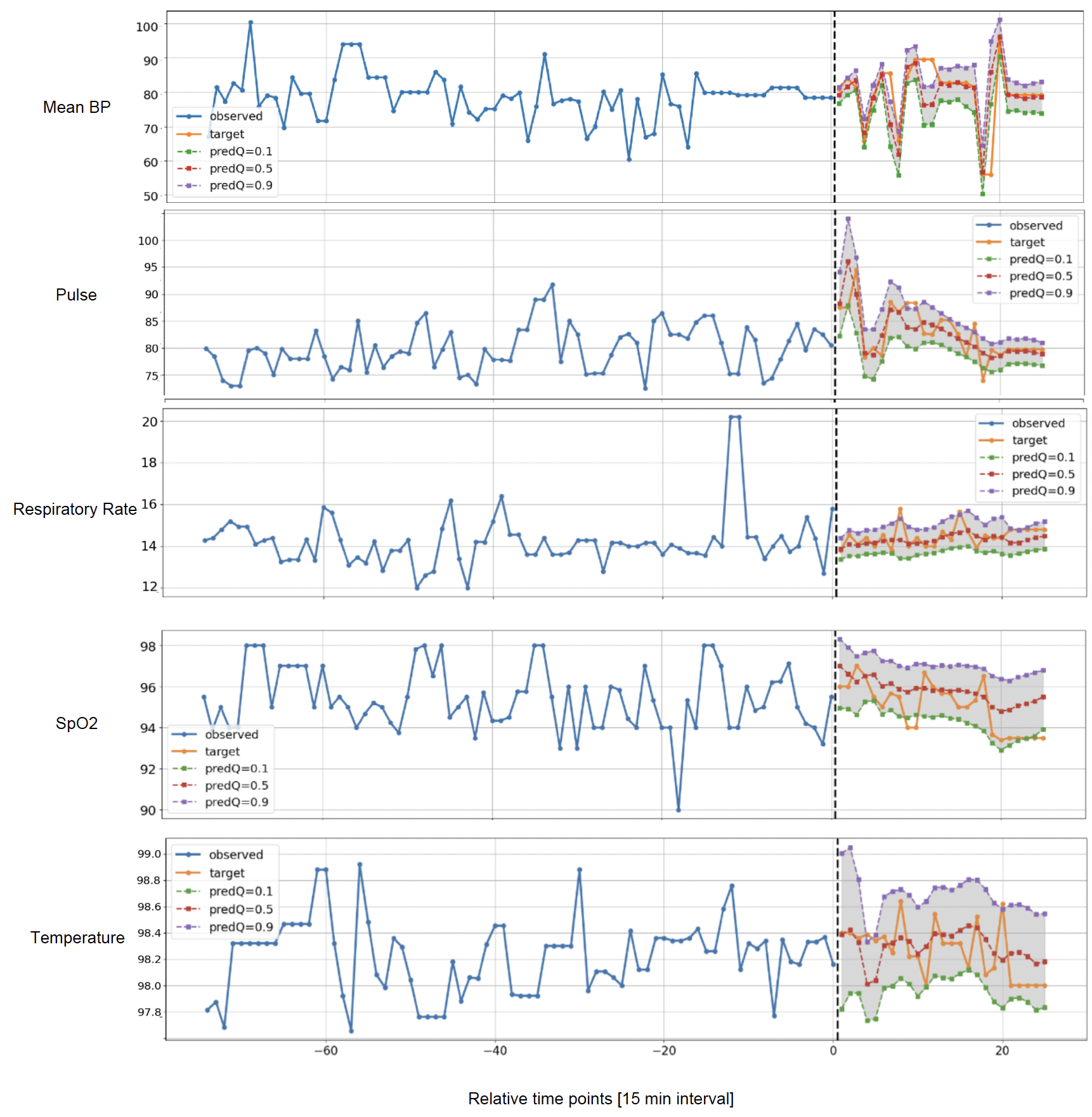}}
\caption{Sample prediction visualization across vital features. Different colors indicate different trajectories: blue indicates observed historical data; orange indicates ground truth; green, red and purple indicate the 10th, 50th and 90th percentile prediction respectively.}
\label{trajectory}
\end{figure}

In Fig. \ref{trajectory}, we show a visualization of a sample output for a subject in the external validation set. For mean BP and pulse, the 50th percentile follows closely to the true values and fluctuations. In vitals with more missingness such as SpO2 and temperature, TFT-multi estimates a reasonable bound but does not track fluctuations well, which is concordant with our findings above. 

\begin{figure}[h!]
\centerline{\includegraphics[width=0.5\columnwidth]{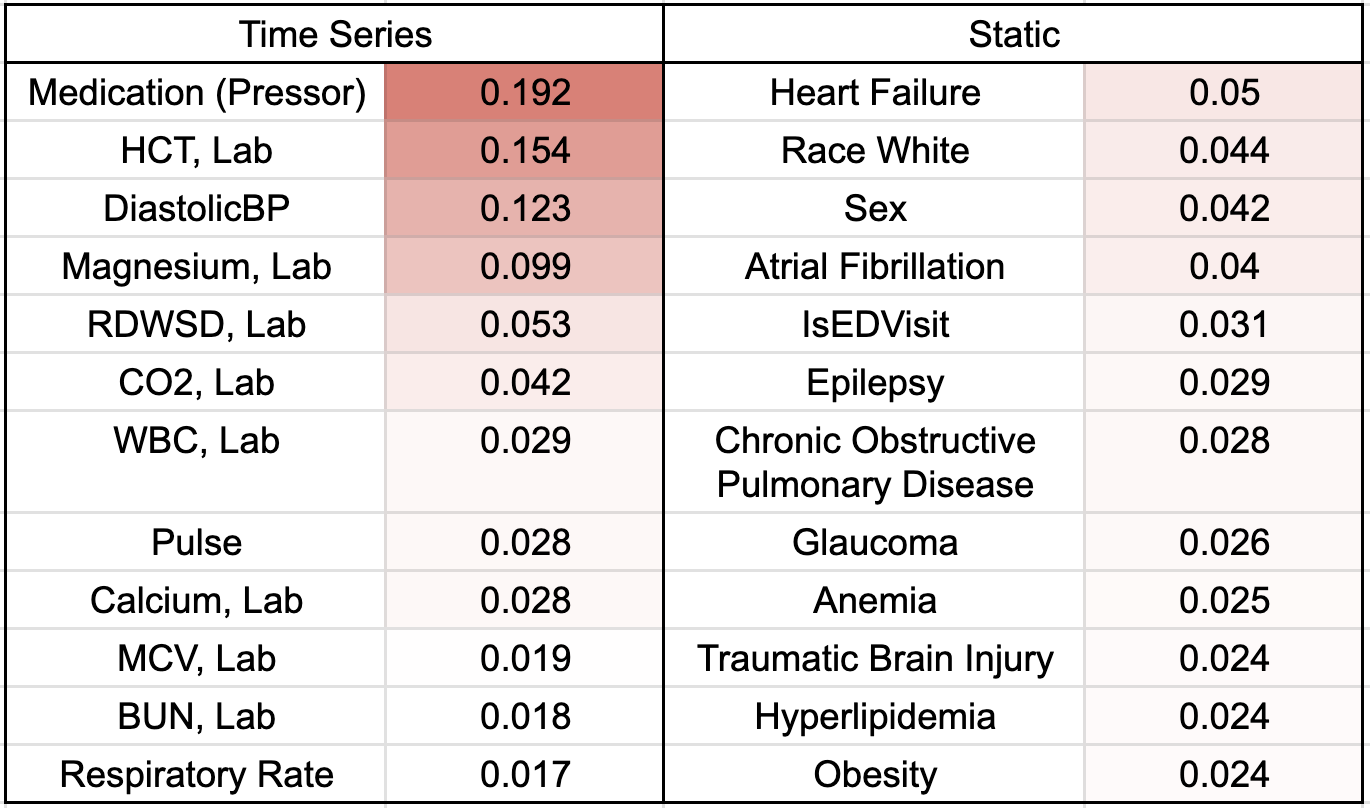}}
\caption{Top features by importance in predicting the 50th percentile trajectory for both static and time series variables. Color gradient reflects importance by weight.}
\label{import}
\end{figure}

\subsubsection*{Impact analysis of available historical data on performance}
To evaluate the effect of the length of historical data provided on predictions, we run additional experiments by comparing model performance between 3 ``lookback'' window lengths: 3, 9 and 18.75 hours, and compare results in Table \ref{lookback}. While the three options performed comparably, a larger ``lookback'' window leads to better predictive performance in most measures in the testing set and all measures in the validation set. However, even with the smallest window size (3-hr), TFT-multi outperforms other methods in Table \ref{MAE} that had more historically available data in measures with a large missing rate, confirming that a simultaneous prediction model will outperform several individual models, especially in vitals with large missingness.

\begin{table}[ht]
\centering
\begin{tabular}{|>{\centering\arraybackslash}m{3cm}|>{\centering\arraybackslash}m{2cm}|>{\centering\arraybackslash}m{2cm}|>{\centering\arraybackslash}m{2cm}|>{\centering\arraybackslash}m{2cm}|>{\centering\arraybackslash}m{2cm}|}
\hline
Window length & Mean BP & Pulse & SpO2 & Resp & Temp \\
 \hline
3-hr (test) & 10.1 [11.1] & 8.6 [10] & 1.8 [2.1] & 2.7 [13.9] & 0.8 [0.8]\\
\hline
9-hr &  8.6 [9.8] & \textbf{8 [9.8]} & 1.8 [2.1] & \textbf{2.6 [13.9]} & 0.8 [0.8] \\
\hline
18.75-hr & \textbf{7.4 [8.1]} & 9 [8.9] & \textbf{1.8 [2]} & 4.5 [13] & \textbf{0.7 [1]} \\
\hline
\hline
3-hr (EV) & 10 [11.8] & 7.6 [9] & 2.3 [2.5] & 3.9 [24.4] & 0.85 [0.8] \\
\hline
9-hr & 9.9 [11.4] & 7.5 [9.2] & 2.7 [2.8] & 3.4 [24.4] & 0.95 [1] \\
\hline
18.75-hr & \textbf{6.2 [8.9]} & \textbf{7.2 [7.8]} & \textbf{1.8 [2.8]} & \textbf{2.4 [30.1]} & \textbf{0.8 [0.7]} \\
\hline
\end{tabular}
\caption{Model performance across vital signs using varying amount of available historical data. Values indicate mean absolute error [mean absolute percentage value(\%)] for held-out test and external validation set. In general, more historical data achieves better performance (highlighted in bold).}
\label{lookback}
\end{table}

\subsection*{Prediction bound estimation}
\begin{figure*}[h]
\centering
    \begin{subfigure}[b]{0.8\textwidth}
        \centering
        \includegraphics[width=0.8\textwidth]{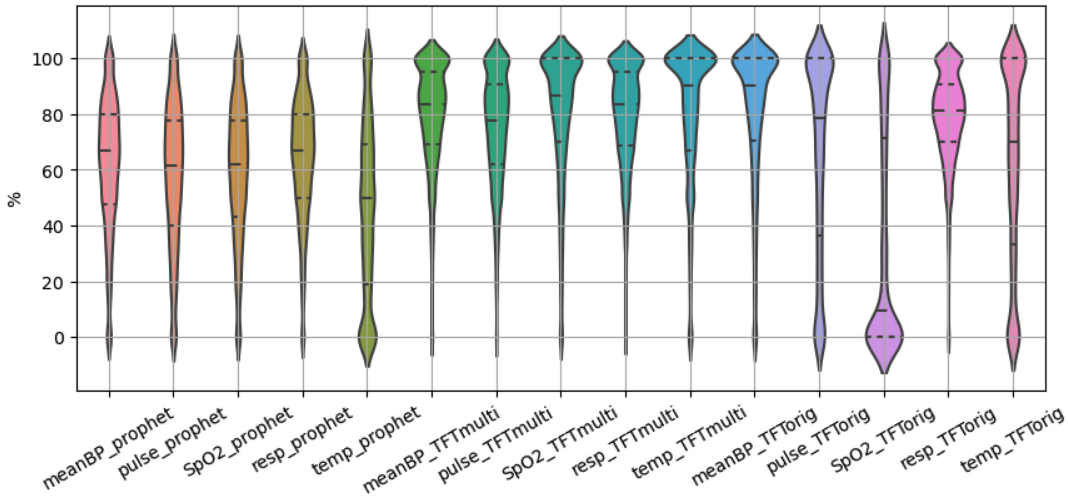}
        \caption{ }
        \label{violin_test}
    \end{subfigure}
    
    \begin{subfigure}[b]{0.8\textwidth}
        \centering
        \includegraphics[width=0.8\textwidth]{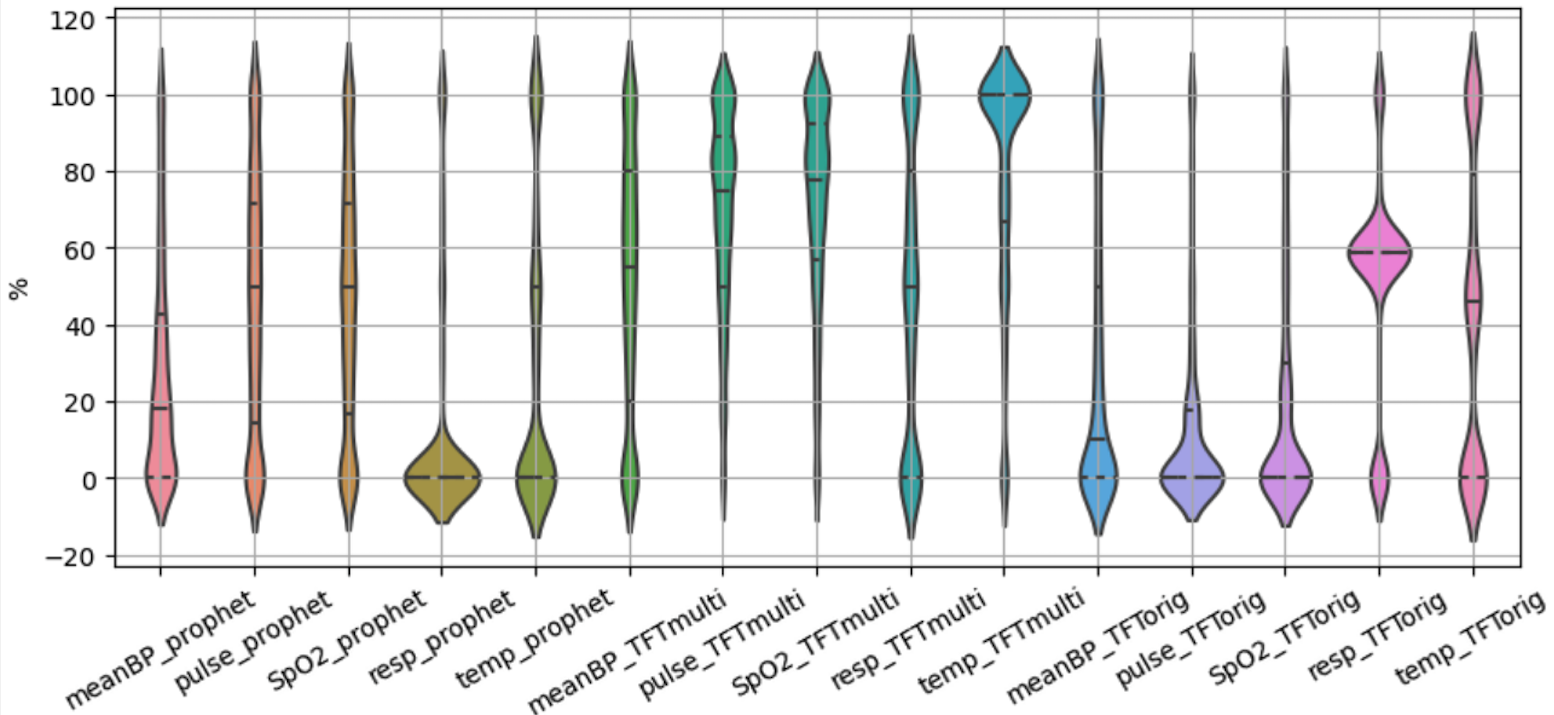}
        \caption{ }
        \label{violin_mimic}
    \end{subfigure}
    \caption{Violin plot comparing percentage of true trajectory within the estimated bounds across vitals of interest for three methods: prophet, TFT-multi (ours) and TFT. Lines within each plot represent the first, second (median) and third quartile values of the held-out test set in (a) and external validation set in (b).}
    \label{violin}
\end{figure*}

In clinical support systems, prediction bounds are sometimes more helpful than singular values, as vital measures have high interpersonal differences and a value is considered abnormal only if it falls outside a range. Therefore, we perform an additional analysis to better understand model performances in predicting upper and lower bounds. For each subject in both validation sets, we calculate the percentage of times a true value is within prediction bounds. {For Prophet, we use y\_hat\_upper and y\_hat\_lower as bounds, which capture uncertainty in trend, seasonality and observation noise in the prediction\cite{taylor2018forecasting}.} For TFT\cite{lim2021temporal} and TFT-multi, we use the 10th and 90th quantile predictions. Since VAR\cite{zivot2006vector} does not offer any bound, it is excluded in this analysis. In Fig. \ref{violin_test}, we show the percentage distribution over the test set for all 5 vital signs in a violin plot and the first, second and third quartile values as lines within each plot. Across all vitals, TFT-multi correctly bounds the most number of points within the entire trajectory. Consistent with the Bland-Altman analysis (Fig. \ref{bland}), our findings suggest that for values that have high missingness such as SpO2, metric calculation using only real values may not be the best as there are few values. In Fig. \ref{violin_mimic}, we show results for the external set, where performance across methods have decreased as expected, but TFT-multi still remains the best performing method. 

\subsection*{Feature importance}\label{feat_importance_section}
We conduct a feature importance test, as proposed in the original TFT paper\cite{lim2021temporal}, by aggregating the multi-head attention weights for each input feature to see which input had the most influence over the predictions. We show the top features by weight for both the time series and static inputs in Fig. \ref{import}. For time series inputs, medication has the most influence over vital predictions. This aligns with our expectation, as pressors should influence blood pressure significantly within a relative short period of time. Among static variables, comorbidities such as heart failure, arterial fibrillation and chronic obstructive pulmonary disease have the most effects over vital signs. We note here the feature importance is reported across all patients and time points, which may not be informative on a subject level and is a shortcoming of this analysis. For each subject, the factors that affect their prediction most may be different and will be studied in future works.


\subsection*{Treatment response forecasting}
We extend our model's application beyond forecasting and conduct a study case in medication effect estimation. For patients who received pressor administrations within the course of their ICU care, we forecast three trajectories: (1) assuming the true administration; (2) assuming continuous administration (all 1's); and (3) assuming no administration (all 0's). First, we calculate a MAE for real points between these scenarios and report the results: 7.44 when the ground truth pattern is used, 7.47 when setting pressor administration to 1, and 7.46 when setting pressor administration to 0. As expected, the actual medication input pattern produces predictions most closely fit with observation than setting the medication to all 1s or 0s. Second, we apply t-tests to test that the modeled effect of pressor administration was consistent with its intended effect. We present our t-test results in Table \ref{t-test}, where for $\hat{y}$ we use the 50th percentile prediction. We perform 3 comparison tests, the first in which we compare the mean trajectory prediction between the two hypothetical scenarios without considering the ground truth. In row 1, we see an average increase of 3.67 in blood pressure when assuming some medication is administered. The second and third comparisons take real observations into account and test for predicted differences between scenarios for time points that have a ground truth observation (obs) of 1 or 0. When the ground truth observation is no medication (obs=0), hypothetically modeling pressor administration predicts a value 3.12 higher than the alternative. When the patient was given medication in real life, the difference is 2.19. In all three scenarios, the p-values are significant ($<0.05$) to reject the null hypothesis and could indicate that our model is learning the association between increased BP and pressor administration.

\begin{table}[ht]
\centering
\begin{tabular}{|>{\centering\arraybackslash}m{10cm}|>{\centering\arraybackslash}m{2cm}|>{\centering\arraybackslash}m{2cm}|}
\hline
Hypothesis & Statistic & P-value \\
 \hline
mean($\hat{y}|$medication=1) - mean($\hat{y}|$medication=0) = 0 & 3.67 & 2e-4 \\
\hline
mean($\hat{y}|$medication=1, obs=1) - mean($\hat{y}|$medication=0, obs=1) = 0 & 3.12 & 1e-3 \\
\hline
mean($\hat{y}|$medication=1, obs=0) - mean($\hat{y}|$medication=0, obs=0) = 0 & 2.19 & 0.028 \\
\hline
\end{tabular}
\caption{T-test trajectory comparison for forecasting mean blood pressure conditioned on medication (pressor) intake.}
\label{t-test}
\end{table}


\section*{Discussion}

In this work, we extend TFT, a multi-horizon time series prediction tool, from univariate to multivariate and develop a novel end-to-end pipeline to simultaneously predict 5 vital signs for patients in the ICU. We compare our method against state-of-the-art methods and show our model's competitive performance across vital signs in both the test set and external validation set. While the original TFT may have difficulty predicting variable trajectories with high missingness such as SpO2 by itself, TFT-multi uses its relation with other vitals such as pulse to improve prediction. Though we observed limited calibration of our model in predicting respiratory rate, SpO2 and temperature, we hypothesize a contributing factor to be due to the data itself. Respiratory rate has a lower sampling rate and may not provide enough data for the model to fully learn its pattern (as observed in Fig. \ref{bland_resp}); SpO2 and temperature have a limited range of observations in our training data, as observed in Fig. \ref{bland_SpO2} and \ref{bland_temp}. In addition to its competitive performance in predicting trajectories, we highlight our model's advantage in requiring much less computational resources than state-of-the-art univariate models, where they require training multiple models to predict multiple features, our model can predict them all in one run. Lastly, we demonstrate a use case of our pipeline for treatment effect estimation, which can be extended into a counterfactual estimation tool as a next step.

Next, we discuss some limitations and future directions. First, the model has limited calibration in predicting features with high missingness and have difficulty tracking frequent fluctuations in some cases. Second, interpretability in generative models has been a long-standing issue\cite{amann2020explainability}. While we try to interpret the model by analyzing aggregated attention weights, further work can be done to analyze feature importance on a subject-level. Lastly, in our proof-of-concept medication analysis, we assumed a binary administration of pressors while in reality medications are given intermittently and at varying doses. Future work will include modeling these medications inputs with more variability to test performance in a more realistic setting.

With many future directions, our pipeline has the potential to be incorporated into clinical practice in several ways. First, it can be extended into a real time early warning system, which can assist clinicians in monitoring ICU patients and predict anomalous events up to 6 hours prior. Second, the model can be extended to include other data modalities including medical images and text data, which are highly complex and often serial in nature. Another direction for this work is to extend it to the outpatient cohort, which is more representative of the population but has large amounts of missing in time series data. A model to first impute irregularly sampled time series data can be added in the pipeline to address this issue and predicts longer range trajectories in the outpatient cohort, benefiting a larger population. 


\section*{Methods}
\subsection*{Data sources}
This study was performed using two independent datasets. Model development leveraged the publicly available MIMIC-IV v2.2\cite{johnson2020mimic, johnson2023mimic} dataset. MIMIC-IV is an EHR dataset consisting of patients presenting at the Beth Israel Deaconess Medical Center. De-identified electronic health records were also extracted from a large academic medical center for model validation. The study design consisted of retrospective analysis using de-identified data and was thus deemed non human-subjects research by the local IRB. Informed consent and ethics approval are waived by the approval of local IRB. The study was conducted in accordance with relevant guidelines and regulations. 

\subsection*{Data preparation}
We train and validate our model using the MIMIC dataset and perform external validation on the second private institutional dataset. 
After initial filtering, the MIMIC set consists of 9,638 patients who were admitted to the intensive care unit (ICU). The institutional cohort consists of 35,286 patients who presented at the Emergency Department and were eventually admitted to the ICU. We summarize some key characteristics of both cohorts in Table \ref{table1}.

\begin{table}[ht]
\centering
\begin{tabular}{|>{\raggedright\arraybackslash}m{5cm}|>{\centering\arraybackslash}m{4cm}|>{\centering\arraybackslash}m{4cm}|}
\hline
\textbf{Sample characteristics} & \textbf{MIMIC (n=9638)} & \textbf{Institutional (n=35286)} \\
\hline
length of stay (days) & 3.57$\pm$0.25 & 14$\pm$22 \\
\hline
\multicolumn{3}{|l|}{\textbf{Demographics}} \\
\hline
Male (\%) & 45.6 & 43.2 \\
\hline
Age & 62$\pm$16 & 55$\pm$23 \\
\hline
\multicolumn{3}{|l|}{\textbf{Ethnicity (\%)}} \\
\hline
Caucasian & 68 & 42 \\
\hline
African American & 10 & 11 \\
\hline
Asian, Pacific Islander & 3.1 & 10 \\
\hline
Native American & 0.2 & 0.2 \\
\hline
\multicolumn{3}{|l|}{\textbf{Comorbidities (\%)}} \\
\hline
Asthma & 13.1 & 6.7\\
\hline
Diabetes & 34.3 & 17.8 \\
\hline
Heart Failure & 32.1 & 7.6\\
\hline
\multicolumn{3}{|l|}{\textbf{Medication (\%)}} \\
\hline
Pressors & 30 & 30 \\
\hline
\end{tabular}
\caption{Cohort characteristics of MIMIC (train/test) and private institutional (external validation) dataset, summarized by mean$\pm$standard deviation or percentage indicated by $\%$.}
\label{table1}
\end{table}

We include static and time series, numerical and binary variables including 5 demographic variables such as age and BMI, 72 comorbidities including heart failure and diabetes, 21 laboratory results including blood glucose and potassium, 12 medications in the class of pressors including Epinephrine and Angiotensin, and 5 common vital signs: mean arterial blood pressure (mean BP), pulse, SpO2, respiratory rate (Resp) and temperature (Temp). We resample all time series data into 15 minute intervals and aggregate using the mean (for numeric variables) or median (for categorical variables) within each interval, and then forward-fill for any missing intervals (similar to past approaches \cite{jawadi2024predicting}). For static variables, we fill missing values with 0 for binary inputs such as comorbidities and drop samples with missing numeric values. For medications, we group all pressor administrations into one categorical variable and assume a missing value indicates no medication given. For each encounter, we construct 100 time points and use the first 75 as the past (18.75 hours) and the last 25 as the future (6.25 hours). During model training, different variables are considered as follows: demographics and comorbidities as past static variables; laboratory results, vital signs, age and medications up until the first 75 time points as past time series variables; age in the last 25 time points as known future time series variables. We randomly split the cohort to 80\% training set and 20\% held-out test set.

\subsection*{Temporal fusion transformer}
In this section, we briefly introduce temporal fusion transformer (TFT)\cite{lim2021temporal}, a transformer based model for multi-horizon time series forecasting. TFT introduces a novel way to incorporate static and time series covariates separately, preserving both short and long range dependencies between variables. There are several building blocks that make TFT powerful in time series prediction. First, historical time series inputs are passed through sequence-to-sequence layers to generate context vectors for short range dependencies. Meanwhile, static covariates are passed through an encoder to generate a representation, which is then combined with the context vectors in the static enrichment layer. Next, these embeddings are fed into multi-head attention blocks, which are better at capturing long term relations\cite{vaswani2017attention}. Lastly, the output from the attention layer is processed by more densely connected layers to produce prediction forecasts. In multiple parts of the architecture, variable selection units are incorporated to select relevant covariates to pass onto the next layer. TFT allows different data types to be incorporated at different points in the model: past static variables, past time series variables, and future time series variables. The model outputs multi-horizon forecasts for 3 quantiles: the 10th, 50th and 90th. While the 50th quantile is often closest to the ground truth, the 10th and 90th quantiles act as prediction upper and lower bounds. By incorporating and processing different types of covariates and combining strengths from short and long term dependencies, TFT has shown promising results in predicting electricity, traffic and stock market patterns\cite{lim2021temporal}. In addition, feature weights can be extracted across multi-head attention layers for interpretability. In other studies, TFT have shown comparable results in vital prediction\cite{bhatti2024vital}, prompting our work to extend it to simultaneous prediction for multiple variables in sparsely sampled data.

\begin{figure}[!t]
\centerline{\includegraphics[width=0.5\columnwidth]{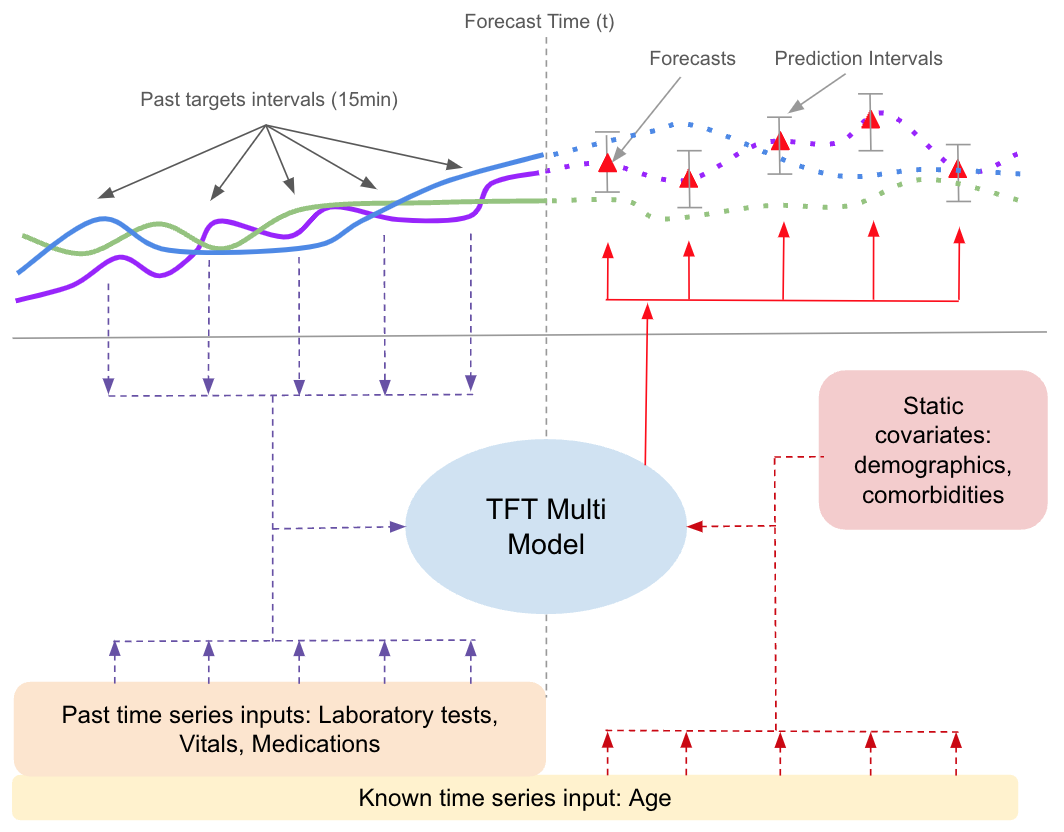}}
\caption{Model workflow for simultaneously predicting 3 example variables.}
\label{in-out-illust}
\end{figure}

\subsection*{TFT-multi}
In this section, we propose TFT-multi, our extension to the original model, that is capable of handling sparsely sampled data and simultaneously predicting multiple measures of interest (in this work 5 vital signs in ICU monitoring). In Fig. \ref{in-out-illust}, we show an illustration of our end-to-end pipeline that takes in time series and static covariates and simultaneously predicts 3 toy trajectories. 
We modify TFT in two main ways: the input-output structure and the loss function, while keeping the inner network the same. In the original model, past target values are appended to the past time series covariates, which we now extend from one to multiple at each time point. In the output, instead of predicting three quantiles at each time point for one target, we modify the last fully connected layers to output three quantiles for each target variable all at once. 

We make two modifications in the loss function. First, we sum over trajectory losses over all quantiles and variables to jointly optimize all target variables. Second, we apply a masking technique to only account for losses in the real values and not the imputed ones to avoid overfitting to imputed data in sparsely sampled variables and increase predictive performance. Our loss function is as follows:

\begin{equation}
\mathcal{L}(\Omega) = \sum_{v \in V} \sum_{y_i \in \Omega} \sum_{q \in \mathbb{Q}} \frac{\sum_{t = 1}^{t_{\text{max}}} QL(y_{it}^v, \hat{y}_{it}^v(q), q) \cdot \mathbb{I}_{y_{it}^v}({\text{real value}})}{\sum_{t = 1}^{t_{\text{max}}}\mathbb{I}_{y_{it}^v}({\text{real value}})} 
\end{equation}

\begin{equation}
QL(y, \hat{y}, q) = q * \text{max}(0, (y - \hat{y})) + (1 - q) * \text{max}(0, (\hat{y} - y)),
\end{equation}
where $V$ is the set of target variables, $\Omega$ is the training dataset, $Q$ is the set of quantiles $\{0.1,0.5,0.9\}$ across future time points $t$ to $t_\text{max}$. We use the notation $\mathbb{I}(x)$ as the indicator function with value 1 when $x$ is true and 0 otherwise.

For training the model, we use the Adam\cite{kingma2014adam} optimizer with a learning rate of 1e-3, a batch size of 800 and a dropout rate of 0.3. We train the model for a maximum of 2000 epochs and implement early stopping to obtain the best performing model during training. We implement parallelization across two Tesla V100-PCIE-16GB GPUs, and training the model took 6 hours.

\subsection*{Treatment response forecasting}
We present a study case applying our prediction pipeline to forecast patient trajectories conditioning on whether a patient is administered pressors. Pressors (vasopressors) are a class of drugs which raise blood pressure and are used in the management of hypotension and shock\cite{stampfl2024clinical}. They are clinically indicated when mean blood pressure falls below 60 mmHg. As these drugs are expected to increase blood pressure within a short time, blood pressure forecasts should reflect this trend when provided time-varying medication administration. In this analysis, we use the MIMIC cohort only, as the institutional dataset obtained does not have enough information about the injection time and duration of medications. Our experimental set-up is as follows. First, we train a separate TFT-multi model assuming that medication (pressor) administration is known into the future, i.e. as a future time series variable. Then for each patient in the test set, we make forecasts based on three scenarios: the observed truth for medication input, assuming constant administration throughout the prediction interval (all 1s), and assuming no administration (all 0s). 

\section*{Author contributions statement}
R.Y.H and J.N.C conceived of the experiment, R.Y.H conducted all experiments and analysis, R.Y.H and J.N.C prepared the manuscript. All authors reviewed the manuscript. 

\section*{Data availability statement}
The validation dataset is not publicly available due to institutional policy. The MIMIC training dataset is publicly available and published in \href{https://www.nature.com/articles/s41597-022-01899-x}{paper}, the implementation of TFT-multi is available for reproducibility and can be found at the GitHub repository https://github.com/rosie068/TFT-multi.

\section*{Competing Interests Statement}
The author(s) declare no competing interests.


\end{document}